\documentclass[12pt]{article}
\usepackage{graphicx,cite,amsmath}

\newcommand{\rt}{R (\tau)}

\newcommand{\rhot}{\rho (\tau)}
\newcommand{\rhott}{\widetilde{\rho} (\tau)}
\newcommand{\rhota}{\overline{\rho} (\tau)}
\newcommand{\tp}{\tau_p}
\newcommand{\detp}{|\det (M_p^r - 1) |}
\newcommand{\dett}{|\det (M_\tau - 1) |}
\newcommand{\dettm}{|\det (M_{\tau/m} - 1) |}
\newcommand{\ti}{\tau \rightarrow \infty}

\begin{document}

\begin{center}

{\huge Periodic orbit spectrum in terms of }\\

\

{\huge Ruelle--Pollicott resonances}\\

\vspace{1.0cm}

{\Large P. Leboeuf}
\end{center}

\begin{center}
\noindent {\sl Laboratoire de Physique Th\'eorique
et Mod\`eles Statistiques \footnotemark,\\
\vspace{0.1 cm}
Universit\'e de Paris XI, B\^at. 100, 91405 Orsay Cedex, France} 

\vspace{2 cm}

{\bf Abstract}
\end{center}
Fully chaotic Hamiltonian systems possess an infinite number of classical
solutions which are periodic, e.g. a trajectory ``p'' returns to its initial
conditions after some fixed time $\tp$. Our aim is to investigate the spectrum
$\{ \tau_1, \tau_2, \ldots \}$ of periods of the periodic orbits. An explicit
formula for the density $\rho (\tau) = \sum_p \delta (\tau - \tp)$ is derived
in terms of the eigenvalues of the classical evolution operator. The density
is naturally decomposed into a smooth part plus an interferent sum over
oscillatory terms. The frequencies of the oscillatory terms are given by the
imaginary part of the complex eigenvalues (Ruelle--Pollicott resonances). For
large periods, corrections to the well--known exponential growth of the smooth
part of the density are obtained. An alternative formula for $\rho (\tau)$ in
terms of the zeros and poles of the Ruelle zeta function is also discussed.
The results are illustrated with the geodesic motion in billiards of constant
negative curvature. Connections with the statistical properties of the
corresponding quantum eigenvalues, random matrix theory and discrete maps are
also considered. In particular, a random matrix conjecture is proposed for the
eigenvalues of the classical evolution operator of chaotic billiards.

\vspace{1.5cm}

\noindent PACS numbers: 05.45.-a, 05.45.Mt \hfill\break

\vspace{1cm}
\begin{math}
\footnotetext[1]{Unit\'e Mixte de Recherche de l'Universit\'e de Paris XI et
du CNRS}
\end{math}

\newpage

\section{Introduction} \label{1}

In chaotic Hamiltonian systems, the unstable classical periodic orbits form a
set of measure zero among all the possible trajectories. However, as has been
emphasized many times, the periodic orbits are of great interest. In
particular, they are very important in the study of the structure of the phase
space dynamics and the transport properties of the motion. Another relevant
aspect of these orbits is their temporal behavior. At a given energy, the
periods $\tp$ of all the periodic orbits $p$ form a discrete set, $\{ \tp \} =
\{ \tau_1, \tau_2, \ldots, \tau_i, \ldots \}$. Many properties of this
sequence are of interest in, e.g., understanding the quantum mechanical
behavior of chaotic systems. For example, semiclassical theories establish a
link between the statistical properties of the quantum eigenvalues and those
of the classical periods $\{ \tp \}$. One is led to answer questions like how
grows the number of periodic orbits with increasing period, or what are the
correlations (if any) between the periods of different orbits. In this
respect, as regards the first question, one of the main results in the field
is the exponential growth of their number with increasing period \cite{pp}.
Concerning the second, there is no definite answer, but semiclassical
arguments based on random matrix theory (RMT) suggest that there exists
correlations between the orbits (orbits with similar period repel each other)
\cite{argaman}.

Our purpose is to further explore the properties of the spectrum of the
periods $\{ \tp \}$ of the periodic orbits of fully chaotic systems. We
restrict the index ``p'' to the primitive orbits only, i.e. repetitions of a
given orbit are excluded. Our main result is an explicit formula that relates
the density of periods of primitive orbits,
\begin{equation} \label{rhop}
\rhot = \sum_p \delta (\tau - \tp) \ ,
\end{equation}
to the eigenvalues of the classical evolution operator, the so--called
Ruelle--Pollicott resonances. These resonances, which generically are defined
by a set of complex numbers (denoted $\{ \gamma \}$), characterize the decay
of correlations in the time evolution of phase space densities \cite{pr}, and
provide important information about the transport properties of the system.
For an introduction see \cite{book,gaspard}.

The set of periods $\{ \tp \}$ can thus be explicitly related to another, more
fundamental, set, the Ruelle--Pollicott resonances $\{ \gamma \}$. In this
way, the properties of the $\tp$'s can directly be related to the properties
of the $\gamma$'s. The results allow, in particular, to make a systematic
analysis of the structure of the spectrum of periods, focusing from the larger
towards the smaller scales.

Semiclassical theories \`a la Gutzwiller or Balian--Bloch \cite{gutz,bb}
describe the density of quantum states in terms of the periodic orbits of the
corresponding classical system. Here, in turn, we establish a connection
between the density of periods of the periodic orbits and the
Ruelle--Pollicott resonances, i.e. the eigenvalues of the classical evolution
operator. In this way, our results allow to establish a link between the
eigenvalues of the quantum and classical evolution operators that, hopefully,
will be useful to more clearly elucidate their properties and correspondences.

Not much is known at present about the distribution in the complex plane of
the eigenvalues of the classical evolution operator of chaotic Hamiltonian
systems (see, e.g., \cite{book}). Paraphrasing Alfredo Ozorio de Almeida
\cite{foot2}, it is fair to say that though of great theoretical interest, the
formula to be derived below merely relates our ignorance of the periodic orbit
spectrum to the even more mysterious maze of the eigenvalues of the classical
evolution operator. However, there is an increasing effort to understand the
properties and physical interpretation of the latter, and their study is a
central theme in several of the most interesting recent developments in
classical chaotic systems, and of their quantum counterparts. For instance, in
quantum systems the Ruelle--Pollicott resonances lurk behind many interesting
effects. They provide the most simple explanation of the long--range non
universal correlations observed in the quantum spectra of bounded Hamiltonian
systems \cite{dahl,aaa}, and show up in experiments that measure the
transmission through open microwave chaotic cavities \cite{sri}. The spectrum
$\{ \gamma \}$ is also a central issue in recent field theoretic approaches
whose aim is to demonstrate the validity of RMT in chaotic systems
\cite{aasa}, and appears in some mathematical models of quantum chaos, the
Riemann zeros \cite{bls2}. More recently, several studies have clarified the
correspondence between the quantum and classical propagators for discrete maps
in the presence of noise \cite{whbms}.

The starting point of our study is an expression of the trace of the evolution
operator as a sum over the periodic orbits (\S \ref{3}). An inversion of that
formula, based on the M\"obius inversion technique, is implemented. Assuming
that the spectrum of the evolution operator consists of isolated resonances,
the inversion leads to a general formula for the density $\rho (\tau)$ in
terms of the eigenvalues $\{ \gamma \}$ (\S \ref{4}). It is also shown, in the
same section, that a natural decomposition of the density emerges, where
resonances located on the real axis determine the average or smooth behavior
of the density, while oscillatory interferent terms arise from the complex
resonances. An alternative and mathematically more accurate formula for
$\rhot$, based on the Ruelle zeta function (instead of the determinant of the
evolution operator), is first presented in \S \ref{2}. It serves as a
reference for the calculations of \S \ref{4}, and complements the results
obtained. Both approaches are compared in \S \ref{4}. Two illustrative
examples are worked out in \S \ref{5}. The first one is the geodesic motion in
a billiard of constant negative curvature. The spectrum of the evolution
operator can be explicitly computed in this case. This allows to write down a
formula for the density of periods of the periodic orbits, thus illustrating
the general approach of \S \ref{4}. The results reproduce, in our formalism,
those of Ref.\cite{as}. The second example is based on the Riemann zeta
function. Though the dynamical basis for this model is hypothetical, it is
included here mainly to stress the existing analogies and similarities with
known results in analytic number theory. Finally, \S \ref{6} contains some
general remarks and conjectures concerning the spectrum of the evolution
operator, inspired by the results of \S \ref{5} and quantum chaos theory.
Special emphasis is put in the connection with the statistical fluctuations of
quantum eigenvalues and random matrix theory. We also show that important
qualitative differences exist between the spectrum of the classical evolution
operator of smooth flows and that of discrete maps.

\section{The density of periods of periodic orbits: Ruelle zeta} \label{2}

In this section the aim is to derive an explicit formula for the density of
periods of the primitive periodic orbits, Eq.(\ref{rhop}). The density will be
expressed in terms of the zeros and poles, located in the complex plane, of a
particular function, the Ruelle zeta function. To simplify, we will from now
on make reference to the zeros and poles of the Ruelle zeta as its
``singularities''. The calculations presented in this section serve as a basis
for those of \S \ref{4}, where the density $\rhot$ will be expressed in terms
of the eigenvalues of the evolution operator.

The starting point is the function
\begin{equation} \label{pts}
P(\tau) = \sum_p \sum_{r=1}^\infty \tp \ \delta (\tau - r \tp) \ .
\end{equation}
The index $r$ accounts for the repetitions (or multiple traversals) of a given
primitive periodic orbit ``p''. The function $P(\tau)$ is naturally associated
with a zeta function. To see this, we reproduce a well known derivation
\cite{book}, which consists in including inside the double sum (\ref{pts}) a
factor $\exp[b (\tau - r\tp)]$, whose value is one because of the delta
function ($b$ is a real positive constant). Then, expressing the delta
function as $\delta (\tau - r \tp) = (2\pi)^{-1} \int_{-\infty}^{\infty}
\exp[i \xi (\tau - r \tp )] d \xi$, and making the change of variables $\xi =
- i \omega$, $P(\tau)$ takes the form
$$
P(\tau) = \frac{{\rm e}^{b \tau}}{2 \pi i} \sum_p \sum_{r=1}^\infty \tp \
{\rm e}^{- b r \tp} \int_{-i \infty}^{i \infty} {\rm e}^{\omega (\tau - r \tp
)} d \omega \ .
$$
For $b$ positive and sufficiently large, the sum is convergent, and can
therefore be interchanged with the integral. Doing that, and making the
additional change of variables $s = \omega + b$, we now get,
$$
P(\tau) = - \frac{1}{2 \pi i} \int_{b -i \infty}^{b +i \infty} d s \
{\rm e}^{s \tau} \frac{\partial}{\partial s} \sum_p \sum_{r=1}^\infty 
\frac{{\rm e}^{- s r \tp}}{r} \ .
$$
Finally, using the expansion $\log (1 - x) = - \sum_{n=1}^\infty x^n /n$,
we obtain the relation,
\begin{equation}\label{pint}
P (\tau) = - \frac{1}{2 \pi i} \int_{b - i \infty}^{b + i \infty} d s \ 
{\rm e}^{s \tau} \frac{\partial}{\partial s} \log Z_R (s) \ ,
\end{equation}
where $Z_R (s)$ is the topological or Ruelle zeta function
\cite{smale,ruelle1},
\begin{equation}\label{ztop}
Z_R (s) = \prod_p \left( 1 - {\rm e}^{- s \tp} \right)^{-1} \ ,
\end{equation}
for ${\rm Re} (s)$ large. 

Equation (\ref{pint}) allows to make a connection between the sum (\ref{pts})
and the analytic properties of $Z_R (s)$. In order to proceed we thus need
some information about the latter. In fact, for certain classes of hyperbolic
systems, it can be shown that $Z_R (s)$ is analytic for ${\rm Re} (s) > h_{\rm
top}$, where $h_{\rm top}$ is the topological entropy, and has a meromorphic
extension to the whole complex plane \cite{ruelle1}. Restricting our analysis
to this case, from (\ref{pint}) an explicit and simple formula follows (with
$b > h_{\rm top}$, and closing the path of integration towards the left part
of the complex plane). It expresses $P(\tau)$ in terms of the location in the
complex plane, denoted $\eta$, of the singularities of $Z_R (s)$,
\begin{equation} \label{ptr}
P(\tau) = \sum_\eta g_\eta \ {\rm e}^{\eta \tau} \ .
\end{equation}
In this equation the integer $g_\eta$ is the multiplicity, and is positive for
poles and negative for zeros. The index $\eta$ has a double significance: it
serves as an index to enumerate the singularities, and also denotes their
location in the complex plane, at $s=\eta$.

Two alternative and distinct formulas for $P (\tau)$ are thus available, one
as a sum over the periodic orbits, Eq.(\ref{pts}), the other as a sum over the
singularities of $Z_R (s)$, Eq.(\ref{ptr}). It is convenient to rewrite
Eq.(\ref{pts}) in terms of the density, using the definition (\ref{rhop}),
\begin{equation}\label{ptrho}
P(\tau) = \tau \sum_{r=1}^\infty \frac{1}{r^2} \ \rho (\tau/r) \ .
\end{equation}
From now on, the idea of the computation is the following. If we manage to
invert equation (\ref{ptrho}), and express the density $\rhot$ in terms of
$P(\tau)$, we are done, because Eq.(\ref{ptr}) can then be used for $P(\tau)$,
and the final output would be an expression of $\rhot$ in terms of the
singularities $\eta$ of the function $Z_R (s)$.

Inversion problems have many important physical applications, but are in
general difficult to solve. In our case, the inversion of Eq.(\ref{ptrho}) will
be based on the M\"obius inversion formula. This is a technique developed in
the nineteenth century, that has been extensively exploited in number theory
\cite{edwards}. More recently, it has found concrete applications in physics,
like for example for computing the phonon density of states from experimental
measurements of the specific heat of solids \cite{chen}.

The inversion proceeds as follows. Consider the sum 
\begin{equation}\label{s}
S_1 = \sum_{m=1}^\infty \frac{\mu (m)}{m} P (\tau/m) \ ,
\end{equation}
where $\mu (m)$ is the M\"obius function \cite{edwards}. This is a
number--theoretic function, whose properties are based on the prime
decomposition of the integer $m$. It is defined as,
$$
\mu (m) = \left\{ \begin{array}{ll}
         1  & \mbox{if $m=1$} \\
    (-1)^k  & \mbox{if $m$ is a product of $k$ distinct primes} \\
         0  & \mbox{if $m$ has one or more repeated prime factors} \ ,
\end{array} \right.
$$
(and thus $\mu (m) = 1,-1,-1,0,-1,+1,\ldots$ for $m=1,2,3,4,5,6,\ldots$, a
quite erratic function). If, in Eq.(\ref{s}), we use for $P$ the series
defined by the r.h.s. of Eq.(\ref{ptrho}), we obtain
$$
S_1 = \tau \sum_{m,r=1}^\infty \frac{\mu (m) \ \rho (\tau/r m)}{(r m)^2}  =
\tau \sum_{n=1}^\infty \sum_{m/n} \frac{\mu (m) \ \rho (\tau/n)}{n^2} = \tau \rhot \ ,
$$
where the sum $\sum_{m/n}$ runs over the divisors $m$ of $n$. The last
equality is the key point of the inversion technique. It follows from a
remarkable property of the M\"obius function, namely $\sum_{m/n} \mu (m) =
\delta_{n,1}$. Finally, combining the last equation with Eq.(\ref{s}), we
obtain an equation for the density,
\begin{equation}\label{rhor2}
\rhot = \frac{1}{\tau} \sum_{m=1}^{m_c} \frac{\mu (m)}{m} P(\tau/m)
      =\frac{1}{\tau} \sum_{m=1}^{m_c} \frac{\mu (m)}{m}
       \ \sum_\eta g_\eta \ {\rm e}^{\eta \tau/m} \ .
\end{equation}
The sum over $m$ has been truncated at a value equal to the integer part of
$\tau/\tau_{min}$, where $\tau_{min}$ is the period of the shortest periodic
orbit of the system,
$$
m_c = \left[ \tau/\tau_{min} \right] \ .
$$
This truncation is a consequence of the fact that $P (\tau) = 0$ for $\tau <
\tau_{min}$ (cf Eq.(\ref{pts})), and therefore $P(\tau/m) = 0$ for $m > m_c$
in Eq.(\ref{rhor2}).

Equation (\ref{rhor2}) is our first result. It expresses the density of
periods of the primitive periodic orbits in terms of the singularities of the
Ruelle zeta function.

The right hand side of Eq.(\ref{rhor2}), should, in principle,
reproduce a series of delta peaks located, according to the definition of
$\rhot$, at the periods of the primitive periodic orbits. In order to better
display the structure of Eq.(\ref{rhor2}), and see the correspondence with the
series of delta peaks, it is useful to consider a decomposition of the density
in two parts,
\begin{equation} \label{rhodec}
\rhot = \rhota + \rhott \ .
\end{equation}
This decomposition is associated to a classification of the resonances into
two distinct sets, each set contributing to one of the two terms in the r.h.s.
of Eq.(\ref{rhodec}). Because $P(\tau)$ is a real function, the Ruelle zeta
satisfies a simple functional equation (which follows from Eq.(\ref{pint})),
$$
\left( Z_R (s) \right)^* = Z_R \left( s^* \right) \ ,
$$
(the star denotes complex conjugate). A singularity $\eta$ of $Z_R (s)$ is
therefore either simple and real, or complex and comes in conjugate pairs
symmetric with respect to the real axis. This property is at the origin of the
decomposition (\ref{rhodec}), with the real singularities contributing to
$\rhota$, and the complex ones to $\rhott$. If the location of the resonances
in the complex plane is written as
$$
\eta = q_\eta \pm i t_\eta \ ,
$$ 
with $q_\eta$ real and $t_\eta$ real positive, then the two contributions to
the density can be expressed as,
\begin{equation}\label{rhotra2}
\rhota = \frac{1}{\tau} \sum_{m=1}^{m_c} \frac{\mu (m)}{m}
\ \sum_{\eta \in \Re} g_\eta \ {\rm e}^{q_\eta \tau/m} \ ,
\end{equation}
and
\begin{equation}\label{rhot2}
\rhott = \frac{2}{\tau} \sum_{m=1}^{m_c} \frac{\mu (m)}{m}
\ \sum_{\eta, \ t_\eta > 0} g_\eta \ {\rm e}^{q_\eta
\tau/m} \cos (t_\eta \tau/m) \ ,
\end{equation}
where in Eq.(\ref{rhotra2}) the sum is made over the real singularities of
$Z_R (s)$, whereas in Eq.(\ref{rhot2}) it is made over the complex ones
located in the upper half part of the complex plane.

The first contribution, $\rhota$, is given by a sum of real exponential terms.
It has therefore a smooth dependence on $\tau$, and describes the average
properties of the density. It reproduces the behavior of the singular sum
(\ref{rhop}) when $\rhot$ is smoothed on a scale which is large compared to
the average spacing between delta peaks. To reproduce the average part of the
density of periods we thus need to know the location of the real
singularities. It is known, in particular, that in hyperbolic systems its
rightmost real singularity, which controls the asymptotic growth when $\ti$,
is a simple pole at $s = h_{\rm top}$ \cite{pp},
\begin{equation} \label{eta0}
\eta_0 = q_0 = h_{\rm top} \ ; \;\;\;\;\;\; g_{\eta_0} = 1 \ .
\end{equation}
The presence of this pole implies, keeping the term $m=1$ in Eq.(\ref{rhotra2}),
\begin{equation}\label{rhoba2}
{\overline \rho}_0 (\tau) \stackrel{\ti}{=} \frac{{\rm e}^{h_{\rm top} \tau}}{\tau} \ .
\end{equation}
We thus recover, to leading order, the well known exponential growth of
periodic orbits in chaotic systems, with the typical rate given by the inverse
of the topological entropy \cite{pp}. However, the inversion procedure
employed here allows to go beyond that result, and obtain subdominant
corrections to the average growth of the density. It is remarkable, indeed,
that the contribution of the pole at $s = h_{\rm top}$ gives also, from
Eq.(\ref{rhotra2}), a series of exponentially large corrections to the
asymptotic behavior, whit signs depending on the M\"obius function,
\begin{equation}\label{rhobf2}
{\overline \rho}_0 (\tau) = \frac{1}{\tau} \sum_{m=1}^{m_c} 
\frac{\mu (m)}{m} \ {\rm e}^{h_{\rm top} \tau/m} 
= \frac{{\rm e}^{h_{\rm top} \tau}}{\tau} - 
\frac{{\rm e}^{h_{\rm top} \tau/2}}{2 \tau} -
\frac{{\rm e}^{h_{\rm top} \tau/3}}{3 \tau} - \ldots \ .
\end{equation}
The first corrections lower the density, while the first positive one occurs
for $m=6$.

Besides the pole at $h_{\rm top}$, other real singularities of $Z_R(s)$, with
real part $q_\eta < h_{\rm top}$, add further sub--leading corrections to the
density of periods. Those located in the negative part of the real axis
contribute with exponentially small corrections. We are not aware of any
generic result about the location in the complex plane of the additional real
singularities, and how their contributions compare to the corrections that
come from $\eta_0$.

The remaining contribution to the density, $\rhott$, behaves quite
differently. This term is responsible for the discrete nature of the spectrum
of periods of the periodic orbits. On top of the average behavior ${\overline
\rho}$, each complex singularity of $Z_R(s)$ adds an oscillatory term to the
density of periods. It is the interference of the oscillatory contributions of
all the complex singularities that, formally, reproduces the delta--peak
structure of Eq.(\ref{rhop}).

The amplitude of each oscillatory term is given by the exponential of the real
part $q_\eta$ of the singularity. In contrast, the inverse of the imaginary
part, $2\pi m t_\eta^{-1}$, is the period of the oscillation. Singularities
with the smaller imaginary part describe long range fluctuations with respect
to the smooth behavior of the density, on scales which may be much larger than
the typical time that separates two neighboring orbits (terms with $m>1$ are
of longer range, but their weight is of lower order in the limit $\ti$). As
singularities with increasing imaginary part are included, details of $\rho
(\tau)$ on smaller scales are resolved. For instance, to distinguish
individual peaks of $\rho (\tau)$ located around $\tau$ requires complex
singularities with imaginary part of the order of the average density of
periods, $t_\eta \sim {\rm e}^{h_{\rm top} \tau}/\tau$.

Eq.(\ref{rhor2}) provides therefore a harmonic decomposition of the density of
periods, were the frequency of each sinusoidal wave is given by a fraction of
the imaginary part of a complex singularity. Since arbitrary high frequencies
are needed to reproduce a delta peak, Eq.(\ref{rhor2}) also implies that,
generically, singularities with arbitrarily large imaginary part should exist.
But we have no additional information about their distribution in the complex
plane.

Eq.(\ref{rhor2}) can be integrated with respect to $\tau$ to obtain a harmonic
formula for the cumulative distribution, or counting function of the periods.
This function is defined as the number of primitive periodic orbits whose
period is smaller than $\tau$,
\begin{equation}\label{ndef}
N(\tau) = \int_0^\tau \rho (x) \ d x = \sum_p \Theta (\tau - \tp) \ ,
\end{equation}
where $\Theta$ is Heaviside's step function. The result of the integration is,
\begin{equation}\label{nr2}
N (\tau) = \sum_{m=1}^{m_c} \frac{\mu (m)}{m} \ \sum_\eta g_\eta \ {\rm Ei}
\left( \frac{\eta \tau}{m} \right) \ ,
\end{equation}
where ${\rm Ei}$ is the exponential integral function. Similarly to
Eq.(\ref{rhor2}), this expression is, formally, exact. The same decomposition
and general remarks as for the density apply to this function.

\section{The trace of the evolution operator} \label{3}

In the previous section, we described the periodic orbit density and the
counting function in terms of the singularities of the Ruelle zeta function.
In this and the next section we will introduce an alternative description of
the density, based on the eigenvalues of the classical evolution operator. Our
motivations for doing this are the following. On the one hand, an explicit
relation between the periodic orbits and the eigenvalues of the evolution
operator is of clear theoretical interest. It helps in understanding the
properties of both sets, as well as their correspondences. On the other hand,
the connection is a central issue in the study of the quantum and classical
behavior of chaotic systems and allows, via semiclassical techniques, to
relate the quantum and classical eigenvalues.

Before presenting in \S \ref{4} the derivation of the alternative description
of the density of periods, we need to previously introduce some definitions
and basic equations related to the classical evolution operator.

Consider a classical system whose dynamical state is defined by the
phase--space coordinate $x$. After a time $\tau$, the point $x$ evolves into a
new state $y = f_\tau (x)$. The kernel of the evolution operator of the
deterministic motion is defined by a Dirac distribution,
\begin{equation} \label{pf}
{\cal L}_\tau (y,x) = \delta (y - f_\tau (x)) \ .
\end{equation}
We shall assume throughout the paper that $f_\tau (x)$ describes the
conservative smooth flow of a closed Hamiltonian system. Open systems are not
treated here, but may be considered as well within the same formalism
\cite{book}.

The trace of the evolution operator is defined as
\begin{equation} \label{rti}
\rt = \int d x \ {\cal L}_\tau (x,x) \ ,
\end{equation}
where $x$ is integrated over the phase space. The function ${\cal L}_\tau
(x,x)$ is the conditional probability density for the system to be at the
point $x$ by the time $\tau$ if the initial state was at the same point. $\rt$
is therefore proportional to the classical return probability at time $\tau$.
The trajectory starting at $x$ and coming back to the same point after a time
$\tau$ defines a closed loop in phase space. Any phase--space closed loop
defines a cyclic motion, because the system returns to its initial conditions.
For a fully chaotic dynamics, an explicit expression of $\rt$ in terms of the
periodic orbits of the system was obtained in Ref.\cite{ce},
\begin{equation} \label{rts}
\rt = \sum_p \sum_{r=1}^\infty \frac{\tp}{\detp} \ \delta (\tau - r \tp) \ .
\end{equation}
The sum is made over the primitive periodic orbits $p$ and over their
repetitions $r$. Each orbit has a period $\tp$ and a monodromy matrix $M_p$.
The latter describes the stability of the orbit. The factor $\detp^{-1}$ is
related to the overlap between an initial cloud, centered initially around the
orbit $p$, and its iterate, after a time $\tau = r \tp$, in a linear
approximation. The function $\rt$ has therefore a peak at the period of each
primitive periodic orbit, or at one of its repetitions, with a weight
inversely proportional to its stability.

Notice that Eq.(\ref{rts}) is similar to Eq.(\ref{pts}), with the important
difference of the stability factor in the denominator in the former. In spite
of this difference, similar steps as those who led in the previous section
from Eq.(\ref{pts}) to Eq.(\ref{pint}) can be followed for $\rt$. They lead to
a connection between the trace of the evolution operator and a zeta function.
We will not repeat them here, but only mention that, for reasons to be
understood below, the sign in the change of variables between $\xi$ and
$\omega$ is the opposite here. The result, analogous to Eq.(\ref{pint}), is,
\begin{equation} \label{rint}
R(\tau) = - \frac{1}{2 \pi i} \int_{a - i \infty}^{a + i \infty} d s \ 
{\rm e}^{- s \tau} \frac{\partial}{\partial s} \log Z (s) \ ,
\end{equation}
where $a$ is a real constant $< 0$. The function $Z(s)$ is called the spectral
determinant, or Smale zeta function, and is defined as \cite{smale,book},
\begin{equation} \label{zs}
Z(s) = \exp \left[ - \sum_{p,r} \frac{{\rm e}^{s r \tp}}{r \detp} \right] \ .
\end{equation}
The complex variable $s$ has units of inverse of time. Eq.(\ref{rint}) relates
the trace of the evolution operator to the analytic structure of the function
$Z(s)$. Under some general conditions valid for a certain class of hyperbolic
systems (and analogous to those assumed in the previous section for $Z_R
(s)$), the Smale zeta function is generically an entire function
\cite{rugh,eckhardt} (i.e., analytic at all finite points in the complex
plane). More complicated analytic structures of $Z(s)$ may arise, like for
example branch cuts that lead to power law decays in intermittent systems
\cite{dahl}. We will ignore here these other possibilities. Assuming,
therefore, that $Z(s)$ is entire, from Eq.(\ref{rint}) the trace of the
evolution operator can be expressed as \cite{ce},
\begin{equation} \label{rtr}
\rt = \sum_\gamma g_\gamma \ {\rm e}^{- \gamma \tau} \ ,
\end{equation}
where the $\gamma$'s are the zeros of $Z(s)$. The complex set of points $\{
\gamma \}$ define the (complex and discrete) spectrum of the evolution
operator. In this equation and in the rest of the paper, the index $\gamma$
has a double significance: it serves as an index to enumerate the zeros, and
also denotes their location in the complex plane, at $s=\gamma$.

The spectrum $\{ \gamma \}$, given by the zeros of the entire function $Z(s)$,
characterizes the relaxation towards equilibrium of classical statistical
ensembles \cite{pr}. The complex zeros are usually called Perron--Frobenius or
Ruelle--Pollicott resonances. We shall, however, often refer to the {\sl
whole} set of zeros (real and complex) as {\sl resonances}.

The positive index $g_\gamma$ in (\ref{rtr}) is the multiplicity of the
resonance. The physically relevant modes being decaying ones, the
corresponding resonances lie in the positive half plane ${\rm Re} (\gamma)
\geq 0$ (this justifies the choice of sign in the derivation of Eq.(\ref{rint})).
It is in general a difficult problem to determine analytically the spectrum
$\gamma$ for a particular system. However, in Hamiltonian systems, where the
energy is conserved, there exists a well defined long--time equilibrium state,
given by the invariant measure on the energy shell with the microcanonical
weight. The existence of this equilibrium state is manifested in the analytic
properties of $Z(s)$ by the presence of a simple ``ergodic'' zero
\cite{gaspard}, located at
\begin{equation} \label{g0}
\gamma_0 = 0 \ , \;\;\;\;\;\;\;\;\;\;\; g_{\gamma_0} = 1 \ .
\end{equation}
The ergodic zero, located at the origin, corresponds to the unique invariant
measure (in the sense of statistical ensembles) in fully chaotic systems. From
Eq.(\ref{rtr}) it follows that this very general property implies that
$\lim_{\ti} \rt \rightarrow 1$, which expresses the equiprobability to find
the system at any phase space point. This fixes the normalization of the trace
of the evolution operator, or ``return probability'' (the volume is set to
one). Other resonances, with ${\rm Re} (\gamma) > 0$, are associated to
decaying modes that describe the process of relaxation of an initial cloud
towards equilibrium.

\section{The density of periods of periodic orbits: Ruelle--Pollicott
resonances} \label{4}

The two expressions of the trace introduced in the previous section,
Eqs.(\ref{rts}) and (\ref{rtr}), relate a sum over the eigenvalues of the
evolution operator to a sum over all the periodic orbits of the system.
Similarly to \S \ref{2}, we can exploit this connection to derive an explicit
formula for the density of periods of primitive periodic orbits, $\rhot$, but
now expressed in terms of the eigenvalues of the evolution operator (instead
of the singularities of the Ruelle zeta function). This formula will take
again the form of a harmonic decomposition, i.e., the density will be
expressed as an interferent sum over oscillatory terms. The frequency of
oscillation of these terms will be given by the imaginary part of the
Ruelle--Pollicott resonances.

To compute $\rhot$, now the starting point is Eq.(\ref{rts}). To express $\rt$
in terms of the density of periods, the main obstacle are the stability
factors, $\detp$. These are not only functions of the period of the orbits,
but depend also on their stability. However, in fully hyperbolic Hamiltonian
systems, where periodic orbits are dense in phase space, the stability or
Lyapounov exponents of long orbits are the same for almost all orbits
\cite{ozorio}. Therefore, for long orbits the stability factor becomes only a
function of the period,
\begin{equation}\label{hav}
\detp \ \stackrel{\ti}{\longrightarrow}  \ |\det (M_{\tp}^r - 1) | \ .
\end{equation}
Within this approximation, and using the definition (\ref{rhop}) of the
density, the function $\rt$ may be rewritten as,
$$
\rt = \tau \sum_{r=1}^\infty \frac{1}{r^2 |\det (M_{\tau/r}^r - 1) |} \ 
\rho (\tau/r) \ .
$$
The factor $|\det (M_{\tau}^r - 1) |$ is in fact a function of the variable $r
\tau$ (cf, for instance, Ref.\cite{book}), and therefore $|\det (M_{\tau/r}^r
- 1) | = \dett$ depends only on the period. The density then takes the form,
\begin{equation}\label{rti3}
\rt = \frac{\tau}{\dett} \sum_{r=1}^\infty \frac{1}{r^2} \ \rho (\tau/r) \ .
\end{equation}

From now on, the situation is similar to that of \S \ref{2}. We should invert
equation (\ref{rti3}), and express the density $\rhot$ in terms of $\rt$.
Then, using Eq.(\ref{rtr}) for $\rt$, we will get an expression of $\rhot$ in
terms of the eigenvalues $\gamma$. It is convenient, to simplify the
computation, to define a new function
\begin{equation}\label{ft}
G (\tau) = \frac{\dett}{\tau} \ \rt = \sum_{r=1}^\infty \frac{1}{r^2} \ 
\rho (\tau/r) \ .
\end{equation}
Then, consider the sum,
$$
S_2 = \sum_{m=1}^\infty \mu (m) G(\tau/m)/m^2 \ ,
$$
where we again make use of the M\"obius function $\mu (m)$. The sum $S_2$ is
evaluated in two different ways. First, using for $G$ the series expansion in
the r.h.s. of Eq.(\ref{ft}). This gives,
$$
S_2 = \sum_{m,r=1}^\infty \frac{\mu (m) \rho (\tau/r m)}{(r m)^2} =
\sum_{n=1}^\infty \sum_{m/n} \frac{\mu (m) \rho (\tau/n)}{n^2} = \rhot \ ,
$$
where the sum $\sum_{m/n}$ is made over the divisors $m$ of $n$. The last
equality follows, as in \S \ref{2}, from the property $\sum_{m/n} \mu (m) =
\delta_{n,1}$. The second way to evaluate $S_2$ is using for $G$ the first
equality in Eq.(\ref{ft}). Combining both, we obtain,
\begin{equation}\label{rhoret}
\rhot = \frac{1}{\tau} \sum_{m=1}^{m_c} \frac{\mu (m)}{m}
\ \dettm \ R(\tau/m) \ .
\end{equation}
As in \S \ref{2}, the sum over $m$ has been truncated at a value equal to the
integer part of $\tau/\tau_{min}$, where $\tau_{min}$ is the period of the
shortest periodic orbit of the system. This truncation is a consequence of the
fact that $R (\tau) = 0$ for $\tau < \tau_{min}$ (cf Eq.(\ref{rts})), and
therefore $R(\tau/m) = 0$ for $m > m_c$ in Eq.(\ref{rhoret}).

Eq.(\ref{rhoret}) gives an explicit connection between the density of periods
and the trace of the evolution operator. To express the density in terms of
the eigenvalues of the evolution operator, we simply use Eq.(\ref{rtr}). This
gives,
\begin{equation}\label{rhor}
\rhot = \frac{1}{\tau} \sum_{m=1}^{m_c} \frac{\mu (m)}{m}
\ \dettm \ \sum_\gamma g_\gamma \ {\rm e}^{-\gamma \tau/m} \ .
\end{equation}

Equation (\ref{rhor}) is our main result. It provides an alternative formula
for the density of periods of primitive periodic orbits. The structure of
Eq.(\ref{rhor}) is very similar to that of Eq.(\ref{rhor2}), with two
important differences: it includes the stability factors, and the sum is made
now over the eigenvalues $\left\{ \gamma \right\}$ of the classical evolution
operator (given by the zeros of $Z(s)$), instead of the singularities of $Z_R
(s)$.

The right hand side of Eq.(\ref{rhor}) should, in principle, also reproduce a
series of delta peaks located, according to the definition of $\rhot$, at the
periods of the primitive periodic orbits. Since the Smale zeta function also
satisfies the functional equation,
$$
\left( Z (s) \right)^* = Z \left( s^* \right) \ ,
$$
the density can be, as in \S \ref{2}, decomposed into two parts, $\rho =
\overline{\rho} + \widetilde{\rho}$, where real eigenvalues of the evolution
operator contribute to $\overline{\rho}$, and complex symmetric ones to
$\widetilde{\rho}$. If the location of the resonances in the complex plane are
written as
$$
\gamma = q_\gamma \pm i t_\gamma \ ,
$$
with $q_\gamma$ and $t_\gamma$ real positive (or zero), then the smooth and
oscillatory contributions to the density can be express as,
\begin{equation}\label{rhotra}
\rhota = \frac{1}{\tau} \sum_{m=1}^{m_c} \frac{\mu (m)}{m}
\ \dettm \ \sum_{\gamma \in \Re^+} g_\gamma \ {\rm e}^{-q_\gamma
\tau/m} \ ,
\end{equation}
and
\begin{equation}\label{rhotrt}
\rhott = \frac{2}{\tau} \sum_{m=1}^{m_c} \frac{\mu (m)}{m}
\ \dettm \ \sum_{\gamma \in \mbox{\scriptsize C}^+} g_\gamma \ {\rm e}^{-q_\gamma
\tau/m} \cos (t_\gamma \tau/m) \ ,
\end{equation}
respectively ($\Re^+$ denotes the positive real axis (including the origin),
and $C^+$ the upper right sector of the complex plane).

How these results compare to Eqs.(\ref{rhotra2}) and (\ref{rhot2})? Let's
start the comparison with the average part. In the limit $\ti$, the leading
order contribution to Eq.(\ref{rhotra}) is given by the real zero with the
smallest real part. That zero is known, it is the ergodic zero denoted
$\gamma_0$ in the previous section, whose location at the origin is the only
generic property known about the spectrum of eigenvalues of the evolution
operator in hyperbolic systems. Keeping only the term $q_\gamma = 0$ and $m=1$
in Eq.(\ref{rhotra}), we obtain
\begin{equation} \label{rhoba}
\rhota \stackrel{\ti}{=} \frac{\dett}{\tau} \ .
\end{equation}
Thus, in the limit $\ti$, the leading order behavior of the average density of
periods is proportional to the stability factor of the orbits. When compared
to the leading order obtained in \S \ref{2} from the leading pole of the
Ruelle zeta function, Eq.(\ref{rhoba2}), this result implies,
\begin{equation} \label{detop}
\dett \stackrel{\ti}{=} {\rm e}^{h_{\rm top} \tau} \ .
\end{equation}
This correspondence is equivalent to the Hannay--Ozorio de Almeida sum rule,
derived in \cite{ho,ozorio} from a uniformity principle. It expresses the
counterbalance between the exponential proliferation of the periodic orbits
and the growth of their (positive) instability. It is also a consequence of
Pesin's equality, that relates the topological entropy to the sum of the
positive Lyapounov exponents \cite{pesin}. Here it has an analytic
significance, it expresses the correspondence between the pole at $h_{\rm
top}$ of the Ruelle zeta function and the ergodic zero at the origin of the
Smale zeta function.

The higher order terms $m>1$ obtain from the ergodic zero in Eq.(\ref{rhotra})
give for the average density a contribution equivalent to Eq.(\ref{rhobf2}).
Besides the ergodic zero, other real resonances ($q_\gamma > 0, t_\gamma = 0)$
of $Z(s)$ add further sub--leading corrections to the density of periods.
However, no generic result concerning their location is known.

The remaining contribution to the density, $\rhott$, gives the oscillatory
terms. In this approach, based on the eigenvalues of the evolution operator,
the amplitude of each oscillatory term is determined by a competition between
the exponential growth of the stability factor $\dettm$ and the exponential
decay of ${\rm e}^{-q_\gamma \tau/m}$. Using the asymptotic approximation
(\ref{detop}) for the determinant, general arguments indicate that the complex
resonances with the smaller real part should be located on the strip $0 <
q_\gamma < h_{\rm top}$, and therefore that the leading oscillatory
contributions have individual amplitudes that grow exponentially in time
\cite{foot1}. The inverse of the complex part, $2\pi m t_\gamma^{-1}$, is the
period of the oscillation. Resonances with the smaller imaginary part describe
long range fluctuations on top of the smooth behavior of the density, on
scales that can be, depending on the value of $t_\gamma$, much larger than the
typical time that separates two neighboring orbits (terms with $m>1$ are of
longer range, but their weight is of lower order in the limit $\ti$). As
resonances with increasing imaginary part are included, details of $\rho
(\tau)$ on smaller scales are resolved. For instance, to distinguish
individual peaks of the density located around a period $\tau$ requires
complex resonances with imaginary part of the order of the average density of
periods, $t_\gamma \sim {\rm e}^{h_{\rm top} \tau}/\tau$.

Eq.(\ref{rhor}) therefore provides an alternative harmonic decomposition of
the density, were the frequency of each sinusoidal wave depends on the
imaginary part of the position in the complex plane of the eigenvalues of the
evolution operator (Ruelle--Pollicott resonances). Since arbitrary high
frequencies are needed to reproduce a delta peak, Eq.(\ref{rhor}) implies
that, generically, resonances with arbitrarily large imaginary part should
exist (similarly to the Ruelle zeta function).

Eq.(\ref{rhor}) can be integrated with respect to $\tau$ to obtain an
alternative formula for the cumulative distribution, or counting function of
the periods, Eq.(\ref{ndef}). To compute the integral, an explicit form of the
determinant is needed, that depends on the dimensionality of the system.
However, asymptotically the approximate expression (\ref{detop}) can be used.
This yields,
\begin{equation}\label{nr}
N (\tau) \approx \sum_{m=1}^{m_c} \frac{\mu (m)}{m}
\ \sum_\gamma g_\gamma \ {\rm Ei} \left[ (h_{\rm top}-\gamma) \frac{\tau}{m} 
\right] \ ,
\end{equation}
which has to be compared to Eq.(\ref{nr2}). The same decomposition and general
remarks as for the density apply for this function. The main contributions to
the smooth part are obtained from the ergodic zero, $\gamma = 0$, and
oscillatory terms are given by the complex resonances.

We have therefore obtained two alternative and distinct descriptions of the
density of periods of the periodic orbits. From a mathematical point of view,
Eq.(\ref{rhor2}) is preferable to Eq.(\ref{rhor}), because in the derivation
of the latter we have ignored some possible fluctuations of the stability
factors that may occur at short times. However, from a physical point of view
the latter description is clearly more interesting, since the
Ruelle--Pollicott resonances, or eigenvalues of the classical evolution
operator, have a clear and intrinsic physical content, directly related to the
classical dynamics. In practice, it is interesting to exploit both approaches.
This leads to consider in more detail the relationships between them.

As we have shown, the ergodic zero of $Z(s)$ and the pole of $Z_R (s)$ at
$h_{\rm top}$ carry similar (though not exactly equivalent) information. The
pole at $h_{\rm top}$ leads to the smooth contribution (\ref{rhobf2}), whereas
the ergodic zero generates a series which is identical but with ${\rm
e}^{h_{\rm top}\tau/m}$ replaced by $\dettm$. It is only in the asymptotic
approximation (\ref{detop}) of the stability factor that their contributions
coincide. It is interesting to explore the correspondence between the analytic
structure of both functions in the latter approximation. The simplest
procedure is to make, in Eq.(\ref{zs}), the replacement $\detp \approx {\rm
e}^{h_{\rm top} r \tp}$. This approximate Smale zeta function takes the form,
\begin{equation}\label{zstop}
Z(s) \approx \exp \left[ - \sum_{p,r} \frac{{\rm e}^{(s-h_{\rm top})r \tp}}{r}
\right] = \prod_p \left[ 1 - {\rm e}^{(s-h_{\rm top}) \tp} \right] =
Z_R^{-1} (h_{\rm top} - s) \ ,
\end{equation}
where we have used the expansion $\log (1 - x) = - \sum_{n=1}^\infty x^n /n$
to compute the sum over the repetitions. In this approximation, the
poles/zeros $\eta$ of $Z_R (s)$ are mapped into zeros/poles $\gamma$ of
$Z(s)$, located at $\gamma=h_{\rm top} - \eta$. This correspondence is
expected to hold for long times, where the approximation of the stability
factor made before is valid. Using the location of the singularities of $Z_R
(s)$, the mapping (\ref{zstop}) should therefore accurately describe the
location in the complex plane of the zeros of $Z(s)$ that are responsible for
the asymptotic behavior of the density. Those are the zeros whose real part is
small. It maps, in particular, the pole of $Z_R (s)$ at $h_{\rm top}$ into the
ergodic zero of $Z(s)$ at the origin. In contrast, resonances with real part
far away from the origin need not be in correspondence. Moreover, the
approximation (\ref{zstop}) produces a function $Z(s)$ which is meromorphic,
rather than entire, as assumed before. Globally, the mapping (\ref{zstop}) is
clearly false.

\section{Illustrative examples} \label{5}

The aim now is to find systems were the analytic structure of $Z (s)$ and $Z_R
(s)$ can be computed, thus allowing to write--down explicitly the formula for
the density of periods of the periodic orbits, and to compare and illustrate
both approaches. Another interest of such a study is to gain some insight into
the distribution of the resonances and singularities in the complex plane in
concrete examples. The explicit computation of the analytic structure of the
zeta functions is a quite difficult task in general. However, it is doable in
some cases. Two examples are treated in detail. The first one is the geodesic
motion on a two--dimensional manifold of constant negative curvature. The
second, a mathematical model, is based on Riemann's zeta function. Both have
their own peculiarities: the somewhat unphysical character of the motion in
the first, the purely speculative dynamical interpretation in the second. In
spite of them, they both provide a concrete illustration of the general
results obtained previously. The geodesic motion on a space of negative
curvature has long served as a paradigm of classical and quantum chaotic
motion \cite{gutz,bv}.

\subsection{Geodesic flow on surfaces of constant negative curvature} \label{5a}

We are interested in the geodesic motion on a two--dimensional hyperbolic
geometry. In particular, we consider billiards on surfaces of constant
negative curvature, the so--called Hadamard--Gutzwiller model. We will not
enter into a detailed description of the classical and quantum motion in such
surfaces, and refer the reader to the excellent introductory articles
\cite{bv}. For these models Gutzwiller's trace formula is exact \cite{gutz2}.
This allows to express the periodic orbit length spectrum in terms of the
quantum eigenvalues \cite{as}. We will here re--derived the formula for the
length spectrum as an illustration of the general formalism based on the
eigenvalues of the evolution operator and singularities of the Ruelle zeta.

Consider a compact and closed surface of area ${\cal A}$ on the
two--dimensional Poincar\'e disk, constructed from a suitable bounded domain on
which appropriate (periodic) boundary conditions have been defined. The
corresponding classical geodesic motion includes a set $p$ of primitive
periodic orbits (closed geodesics on the compact surface). To start with,
consider the corresponding Selberg zeta function \cite{selb},
\begin{equation}\label{zsel}
Z_{S} (s) = \prod_p \prod_{n=0}^\infty
\left( 1 - {\rm e}^{- (s+n) \ell_p } \right) \ .
\end{equation}
The first product is defined over the primitive periodic orbits $p$ of length
$\ell_p$ (in appropriate units). This function differs in its definition from
the Smale and Ruelle zeta functions. $Z_{S} (s)$ is an entire function, with
zeros located at \cite{selb}
\begin{itemize}
\item[a)] $s=1$; $\ g=1$ 
\item[b)] $s=1/2 \pm i p_\alpha$; $\ g=g_\alpha$
\item[c)] $s=0$; $\ g={\cal A}/(2 \pi) + 1$
\item[d)] $s=-k$, $\ k=1, 2, \ldots$; $\ g=(k+1) {\cal A}/(2 \pi)$ \ ,
\end{itemize}
where $p_\alpha = \sqrt{E_\alpha - 1/4}$, $\alpha = 1, 2, \ldots$. The
$E_\alpha$ are the (quantum) eigenvalues of the Laplace--Beltrami operator on
the surface, and the $p_\alpha$ are the corresponding wavenumbers. The
integer $g$ denotes the multiplicity of the zeros. In case b) they depend on
the degeneracy $g_\alpha \geq 1$ of the eigenvalue $E_\alpha$. Because of
topological constraints, ${\cal A}/(4 \pi)$ is a positive integer.

The topological or Ruelle zeta function (\ref{ztop}) is easily obtained from
$Z_{S} (s)$. It is given by \cite{selb},
\begin{equation}\label{ztoph}
Z_R (s) = \prod_p \left( 1 - {\rm e}^{- s \ell_p } \right)^{-1} 
= \frac{Z_{S} (s+1)}{Z_{S} (s)} \ .
\end{equation}
The analytic structure of this meromorphic function directly follows from that
of $Z_{S} (s)$ (see the left part of Fig.1),
\begin{itemize}
\item[a)] pole at $s=1$; $\ g_{\eta}=1$ 
\item[b)] poles at $s=1/2 \pm i p_\alpha$; $\ g_{\eta}=g_\alpha$
\item[c)] pole at $s=0$; $\ g_{\eta}={\cal A}/(2 \pi)$
\item[d)] zeros at $s=-1/2 \pm i p_\alpha$; $\ g_{\eta}=-g_\alpha$
\item[e)] pole at $s=-1$; $\ g_{\eta}={\cal A}/(2 \pi) - 1$
\item[f)] poles at $s=-k$, $\ k=2, 3, \ldots$; $\ g_{\eta}= {\cal A}/(2 \pi)$.
\end{itemize}
The rightmost pole of this function is real and located at $s=1$. This implies
that $h_{\rm top}=1$ in these systems. This pole is responsible for the
leading asymptotic average growth of the number of orbits, and provides
sub--leading corrections as well. There is an infinite number of other real
poles with $q_\eta < h_{\rm top}$, which also provide sub--leading corrections
to the average part. The complex zeros and poles, aligned here on two
different vertical lines in the complex plane, contribute to the oscillatory
part. There exist poles and zeros with arbitrarily large imaginary part, as
generically required (cf \S \ref{2}).

Taking into account separately the contributions of the real and complex
singularities of $Z_R (s)$, from Eq.(\ref{rhor2}) a formula is obtained for
the length spectrum of the periodic orbits on a compact and closed hyperbolic
surface (with the speed set to one, $\tau$ stands here for the length $\ell$
of the orbits),
\begin{equation} \label{rhodech}
\rhot = \rhota + \rhott \ , \;\;\;\;\;\;\;\;\;\; \tau = {\rm
length~of~orbits,}
\end{equation}
with
\begin{equation}\label{rhoah}
\rhota = \frac{1}{\tau} \sum_{m=1}^{m_c} \frac{\mu (m)}{m}
       \ {\rm e}^{\tau/m} \left[ 1 - {\rm e}^{-2 \tau/m} + \frac{\cal A}{2
       \pi} \frac{{\rm e}^{-\tau/m}}{(1 - {\rm e}^{-\tau/m})} \right] \ ,
\end{equation}
and
\begin{equation}\label{rhoth}
\rhott = \frac{4}{\tau} \sum_{m=1}^{m_c} \frac{\mu (m)}{m}
       \ \sinh (\tau/2 m) \ \sum_\alpha g_\alpha \cos (p_\alpha \tau/m) \ .
\end{equation}

Concerning the complex singularities, as mentioned before $Z_R (s)$ has two
``critical'' lines, one made of complex zeros (located at ${\rm Re} (s) =
-1/2$), the other of complex poles (located at ${\rm Re} (s) = 1/2$). Their
superposition produces the oscillatory part $\rhott$. It is remarkable that
the frequencies of the harmonic decomposition of the periodic orbit density
are directly related to the quantum wavenumbers, a consequence of the fact
that Selberg's trace formula is exact. We will come back to this point later
on. The integrated version of Eq.(\ref{rhodech}) for the counting function, $N
(\tau)$, was computed and analyzed in \cite{as}. For a specific hyperbolic
billiard, we have numerically checked that the deviations with respect to the
leading order behavior ${\overline N} (\tau) \approx {\rm Ei} (\tau)$ observed
in the data were well explained by the sub--leading corrections $m>1$ of the
pole $\eta = h_{\rm top} = 1$ in Eq.(\ref{nr2}).

Notice that each of the terms in the series (\ref{rhoah}) diverges in the
limit $\tau \rightarrow 0$. This divergence is due to the splitting of the
density into two parts, $\overline{\rho}$ and $\widetilde{\rho}$, whereas the
sum of the two terms is well behaved.

Consider now the spectral determinant or Smale zeta function (\ref{zs}), whose
zeros are the eigenvalues of the evolution operator. Another peculiar feature
of the Hadamard--Gutzwiller model is that all the periodic orbits have the
same Lyapounov exponent (equal to one). $M_p$ can be written as a two by two
diagonal matrix whose diagonal elements are ${\rm e}^{\pm \ell_p}$. Therefore,
\begin{equation} \label{deth}
\detp = 2 \ \left[ \cosh (r \ell_p) - 1 \right] \ .
\end{equation}
Expanding the inverse of the determinant as $\detp^{-1} = \sum_{k=1}^\infty \
k \ {\rm e}^{- r k \ell_p}$, and computing the sum over the repetitions $r$,
from (\ref{zs}) and (\ref{deth}) we obtain
\begin{equation}\label{zsh}
Z (s) = \prod_p \prod_{k=1}^\infty
\left[ 1 - {\rm e}^{- (k-s) \ell_p } \right]^{k} \ .
\end{equation}
Using the definition of the Ruelle and Selberg zeta functions,
Eq.(\ref{zsh}) can be re--expressed as
\begin{equation}\label{zsh2}
Z (s) = \prod_{k=1}^\infty Z_R^{-k} (k - s) =
\prod_{k=1}^\infty \frac{Z_{S}^k (k-s)}{Z_{S}^k (k+1-s)} \ .
\end{equation}
Further manipulations of the latter equation lead finally to
\begin{equation}\label{zshf}
Z (s) = \prod_{k=1}^\infty Z_{S} (k-s) \ .
\end{equation}
Since $Z_S (s)$ is entire, it follows from the last expression
that $Z (s)$ is also an entire function. The analytic structure of $Z (s)$
follows from that of $Z_{S} (s)$, with zeros at (cf the right part of Fig.1),
\begin{itemize}
\item[a)] $s=0$; $\ g_\gamma =1$ 
\item[b)] $s=k + 1/2 \pm i p_\alpha$, $k=0,1,2,\ldots$; 
          $g_\gamma = \ g_\alpha$
\item[c)] $s=k$, $\ k=1, 2, \ldots$; $\ g_\gamma= k (k+1) {\cal A}/(4 \pi) + 2$.
\end{itemize}
This distribution of zeros satisfies the ``generic'' requirements concerning
the spectrum of the classical evolution operator of an hyperbolic system: (i)
all zeros have ${\rm Re} (\gamma) \geq 0$, (ii) there is a simple pole at the
origin (ergodic zero), and (iii) there are complex symmetric zeros with
arbitrarily large imaginary part. Using Eq.(\ref{deth}), the density
(\ref{rhor}) is now expressed in terms of the classical resonances as,
\begin{equation}\label{rhorhs}
\rhot = \frac{1}{\tau} \sum_{m=1}^{m_c} \frac{\mu (m)}{m}
\ {\rm e}^{\tau/m} \ \left( 1 - {\rm e}^{-\tau/m} \right)^2 
\ \sum_\gamma g_\gamma \ {\rm e}^{-\gamma \tau/m} \ .
\end{equation}
Using the locations and corresponding multiplicities of the zeros of $Z (s)$
given above, and separating the contributions of the real and complex zeros,
it is easy to check that Eq.(\ref{rhorhs}) is strictly equivalent to
Eqs.(\ref{rhoah}) and (\ref{rhoth}). For the geodesic flow on a constant
negative curvature, the exact density is thus recovered from Eq.(\ref{rhor}),
without any error. The reason for this is that in the present model the
Lyapounov exponents are constant, independent of the orbit. It follows that
Eq.(\ref{hav}) is exact for any period, not only asymptotically (cf
Eq.(\ref{deth})).

\begin{figure}[thb]
\begin{center}
\includegraphics*[width=14cm]{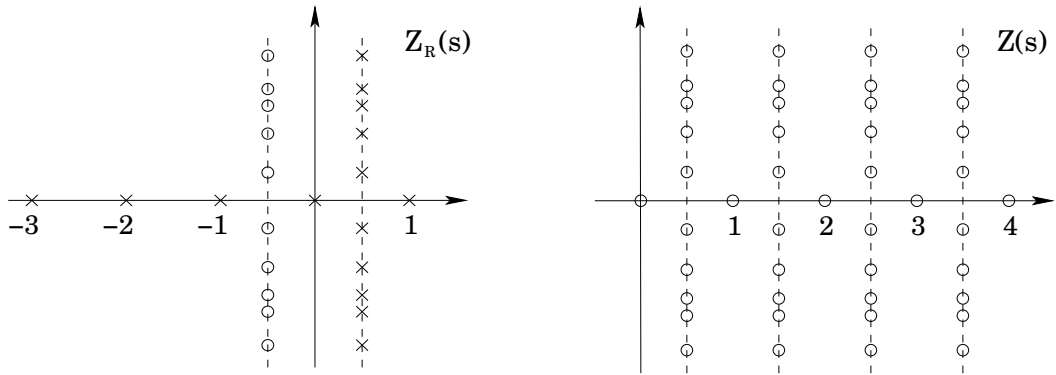}
\end{center}
\caption{\small Analytic structure close to the origin of the Ruelle zeta function
(left) and of the spectral determinant (Ruelle--Pollicott resonances, right)
for a compact and closed billiard on a surface of constant negative curvature.
Crosses are poles, circles are zeros. Multiplicities and exact positions are
given in the text.}
\label{fig1}
\end{figure}

It is however interesting to remark that the two expressions for the density
are obtained from functions with very different analytic structure. In
particular, $Z_R (s)$ is meromorphic and has two ``critical'' lines, one made
of an infinite number of zeros located at ${\rm Re} (s) = -1/2$, the other of
an infinite number of poles located at ${\rm Re} (s) = 1/2$. In contrast, the
spectral determinant is entire and, as complex zeros, has an infinite number
of parallel replicas of the quantum spectrum located at ${\rm Re} (s) = k +
1/2$, $k = 0, 1, 2, \ldots$. The correct result (\ref{rhoth}) for the
oscillatory part of the density is only recovered when the whole set of
complex resonances of $Z(s)$ is taken into account. In contrast, the leading
order asymptotic behavior $\ti$ of the density is controlled by the zero at
the origin and the complex resonances located at ${\rm Re} (s) = 1/2$.

The Hadamard--Gutzwiller model offers also the opportunity to analyze the
accuracy of the approximation (\ref{zstop}) of the Smale zeta $Z(s)$, which
holds asymptotically. It is obtained by keeping only the first term $k=1$ in
Eq.(\ref{zsh2}). In that approximation, the analytic structure of $Z (s)$
consists now of a simple zero at the origin, complex zeros on the line ${\rm
Re} (s) = 1/2$, complex poles on the line ${\rm Re} (s) = 3/2$, and zeros at
$s=k, k=1,2,\ldots$. The analytic structure is thus well reproduced for ${\rm
Re } (s) < 1$ (the zero at the origin and the lowest line of complex zeros
located at ${\rm Re } (s) = 1/2$). The remaining structure is wrong (the
degeneracies of the remaining real zeros is wrong, at ${\rm Re} (s) = 3/2$ it
has a line of complex poles instead of zeros, the other lines of complex zeros
are missing).

\subsection{The Riemann zeta} \label{5b}

Our second example is taken from analytic number theory, which has inspired
several developments in the theory of dynamical systems \cite{ruelle2}. The
results we are going to present are based on a dynamical interpretation of the
Riemann zeros and of the prime numbers. This interpretation is by no means
necessary, but it is a useful one because it introduces the appropriate
physical framework into the discussion, and therefore facilitates the
comparison with dynamical systems. We therefore revisit here some well known
formulas in number theory, viewed from the perspective of the present theory
that connects the periodic orbits to the eigenvalues of the evolution
operator.

The spectral interpretation of the Riemann zeros is based on the following
identification. The imaginary part of each of the Riemann zeros is thought to
be an eigenvalue of a quantum system with a classically chaotic limit with no
time--reversal symmetry. An analysis, based on a semiclassical interpretation
of a formula for the density of the critical zeros \cite{berry}, shows that
the set of prime numbers has to be identified to the set of periodic orbits of
the (unknown) classical dynamics. The analysis leads to the following mapping,
\begin{eqnarray} \label{table}
{\rm primitive~periodic~orbits} \ &\rightarrow& \ {\rm
prime~numbers~} p \nonumber \\
{\rm period~of~the~orbits~~} \tp \ &\rightarrow& \ \log p \\
{\rm stability~} \detp \ &\rightarrow& \ p^r = {\rm e}^{r \log p}  \nonumber \ .
\end{eqnarray}
The last correspondence implies that $h_{\rm top} = 1$ in this hypothetical
dynamical system, that for simplicity we refer to as the Riemann dynamics.

We don't know the classical Hamiltonian behind the Riemann dynamics, but the
information contained in (\ref{table}) concerning the periodic orbits is
enough to write down the trace of the corresponding classical evolution
operator, Eq.(\ref{rts}). Using in the latter the correspondences
(\ref{table}), the trace is expressed as,
\begin{equation} \label{rtsz}
\rt = \sum_p \sum_{r=1}^\infty \frac{\log^2 p}{p^r} \ \delta (\tau - r \log p) \ ,
\end{equation}
where the sum runs over the prime numbers $p$. To write down $\rt$ in terms of
the classical Ruelle--Pollicott resonances, we need to compute the
corresponding spectral determinant, or Smale zeta function. The result,
obtained from (\ref{zs}) and (\ref{table}) is \cite{bls2},
\begin{equation}\label{zsz}
Z (s) = \zeta^{-1} (1 - s) \ ,
\end{equation}
where $\zeta (s) = \sum_{n=1}^\infty n^{-s}$ is the Riemann zeta function.
Contrary to general expectations in bounded hyperbolic systems \cite{rugh},
$Z(s)$ is not entire for the Riemann dynamics, but meromorphic. The analytic
structure of $Z (s)$ follows from that of $\zeta (s)$,
\begin{itemize}
\item[a)] zero at $s=0$, $\ g_\gamma = 1$ 
\item[b)] poles at $s=1/2 \pm i \ t_\alpha$, $\ g_\gamma = - g_\alpha$
\item[c)] poles at $s=1 + 2k$, $\ k=1, 2, \ldots$, $\ g_\gamma = -1$.
\end{itemize}
Assuming the Riemann hypothesis, the $t_\alpha$ are real and define the
position of the $\alpha$'th zero of $\zeta (s)$ along the critical line, of
multiplicity $g_\alpha \geq 1$. The pole of the Riemann zeta transforms into
the ergodic zero of $Z(s)$. The position of the complex Ruelle--Pollicott
resonances coincide here with the complex zeros of the Riemann zeta, but these
are not zeros of $Z(s)$, but poles. Other real poles of $Z (s)$ are generated
by the so--called trivial zeros of the Riemann zeta, now located on the real
positive axis. The global properties of the distribution of the singularities
of $Z(s)$ for the Riemann dynamics satisfy the general requirements of fully
chaotic systems, with the important oddness related to the occurrence of some
poles in place of zeros. 

From (\ref{rtr}) and the analytic structure of $Z(s)$ (assuming $g_\alpha =
1$), the function $\rt$ may be written,
\begin{equation}\label{rtzf}
\rt =  1 - \frac{{\rm e}^{-3 \tau}}{1 - {\rm e}^{-2 \tau}} 
- 2 \ {\rm e}^{- \tau/2} \ \sum_{\alpha=1}^\infty \cos ( t_\alpha \tau ) \ .
\end{equation}
The "anomalous" minus signs that appear in the r.h.s. of this equation
reflect, again, the presence of poles in $Z(s)$. The physical interpretation
of these signs, and of the closely related minus sign that appears in front of
the oscillatory part of the density of the Riemann zeros \cite{berry}, is
unclear for the moment, though an appealing possibility was suggested in
\cite{connes}.

The function $\rt$, here interpreted as the trace of the classical evolution
operator, is in fact a well known function in number theory,
$$
R (\tau = \log x) = \frac{d \psi (x)}{d x} \ ,
$$
where $\psi (x) = \sum_{n\leq x} \Lambda (n)$ ($\Lambda (n)$ is the
Von--Mangoldt function) \cite{edwards}. By inverting Eq.(\ref{rtsz}), as in \S
\ref{4}, and using Eq.\ref{rtzf}), a formula for the density of the logarithm
of the prime numbers is obtained, given by Eq.(\ref{rhor}) with the
appropriate identifications and replacements dictated by the table
(\ref{table}). The result coincides, up to a change of variable, with
Riemann's formula. It can be integrated to obtain the counting function. These
results stress the strong similarities that exist between number theory and
the theory of dynamical systems.

The Ruelle zeta function (\ref{ztop}) coincides with the Riemann zeta
\cite{ruelle2}, $Z_R (s) = \prod_p \left( 1 - p^{- s} \right)^{-1} = \zeta
(s)$. The approximation (\ref{zstop}) is therefore exact in this case, since
Eq.(\ref{detop}) is (cf Eq.(\ref{zsz})).

\section{Concluding remarks} \label{6}

Two different explicit formulas for the density $\rhot$ of periods of the
primitive periodic orbits of fully chaotic classical systems have been
obtained. Both provide a harmonic decomposition of $\rhot$, where the complex
zeros and poles (if any) of the corresponding zeta function are related to the
elementary frequencies of the oscillatory terms, while the real ones
contribute to the smooth part. In one case, the zeta function is the spectral
determinant $Z (s)$. It is assumed to be an entire function; its zeros,
denoted $\gamma$, are the eigenvalues of the classical evolution operator
(usually called Ruelle--Pollicott resonances). In the second formulation, the
relevant function is the Ruelle zeta $Z_R (s)$. In contrast to the spectral
determinant, this function has a meromorphic extension in the complex plane.
The relation between both approaches was discussed in some detail.

The zero of $Z (s)$ located at the origin (or, alternatively, the pole of $Z_R
(s)$ located at $s=h_{\rm top}$) provides the leading average growth of the
density of periods of periodic orbits, $\rhota \stackrel{\ti}{\sim} {\rm
e}^{h_{\rm top} \tau}/\tau$. We found exponentially large sub--dominant
corrections to the leading term that arise from the same zero (or pole), and
that are responsible for the main deviations observed numerically in billiards
in a surface of constant negative curvature.

The zero of $Z (s)$ at the origin reflects a generic property of hyperbolic
systems, the existence of an equilibrium distribution described by the
microcanonical measure. What about the rest of the spectrum? No generic
statements are known, aside the fact that ${\rm Re} (\gamma) \geq 0$, that
resonances with arbitrarily large imaginary part should exist, and that
complex resonances come in symmetric pairs (cf \S \ref{4}). There are,
however, some hints on what could probably be the generic structure, if any,
of the low lying spectrum of $Z (s)$ (e.g., resonances with the smaller real
part), that we would like to discuss now. This low--lying part of the spectrum
describes the long time dynamics of the system. For simplicity, from now on we
restrict the discussion to the case of billiards. We have in mind ``generic''
systems with a discrete spectrum of the classical evolution operator
(exponential decay). We also restrict to ballistic systems (i.e., billiards
whose shape produce a chaotic motion), and are excluding disordered systems.

In \S \ref{5} we saw that for billiards on constant negative curvature the
low--lying spectrum consists of the ergodic zero at $\gamma_0 = 0$, and of an
infinite number of zeros located on the line ${\rm Re} (s) = h_{\rm top}/2$.
This structure also follows for other systems from semiclassical arguments.
For a fully hyperbolic billiard, Gutzwiller's trace formula for the density of
quantum eigenvalues takes the form,
$$
\rho_Q (t) = \sum_\alpha \delta (t - t_\alpha) \approx \overline{\rho}_Q (t) +
\frac{1}{\pi} \sum_{p,r} \frac{\tp}{\sqrt{\detp}} \cos (r t \tp) \ .
$$
As in \S \ref{5a}, $\tp$ denotes here the length of the periodic orbits $p$
and $t_\alpha$ are the (eigen) wavenumbers; $\overline{\rho}_Q (p)$ is the
Weyl term, and we have neglected Maslov indices. Fourier inverting
this formula with respect to $t$, and using the approximation (\ref{detop}),
we obtain the following formula for the density of periods,
\begin{equation}\label{rhosem}
\rhot \approx \frac{{\rm e}^{h_{\rm top} \tau}}{\tau} + \frac{2}{\tau} \
{\rm e}^{h_{\rm top} \tau/2} \ \sum_\alpha \cos (t_\alpha \tau) \ .
\end{equation}
This is precisely the leading order density that is obtained from (\ref{rhor})
assuming that $Z (s)$ has a simple zero at the origin and complex zeros
concentrated on the critical line ${\rm Re} (s) = h_{\rm top}/2$. Thus, the
Hadamard--Gutzwiller model of \S \ref{5} as well as the generalization
Eq.(\ref{rhosem}) suggest that the generic low--lying spectrum of the
classical evolution operator of fully chaotic billiards consists of a simple
zero at the origin plus an infinite sequence (extending to arbitrary large
imaginary parts) of complex symmetric zeros located on the line ${\rm Re} (s)
= h_{\rm top}/2$. A corresponding structure follows for $Z_R (s)$ by
transforming zeros into poles located at $h_{\rm top} - \gamma$. This first
hypothesis about the location of the complex zeros of $Z(s)$ is reminiscent to
that of Riemann in number theory.

For the negative curvature model and in the inverse formula (\ref{rhosem}),
the imaginary part of each Ruelle--Pollicott resonance located on the line
${\rm Re} (s) = h_{\rm top}/2$ coincides with a quantum wavenumber. This
happens because of important non genericities of those systems: the
corresponding semiclassical trace formula (the Selberg trace formula) is
exact, and Maslov indices are zero. If in Eq.(\ref{rhosem}) the Maslov indices
are not neglected (and correction terms are taken into account), the
connection with the quantum wavenumbers is generically lost. The simplest
effect of these phases would be to produce a shuffling of the resonances,
without moving them out of the line ${\rm Re} (s) = h_{\rm top}/2$, and
without changing their statistical properties. Some arguments in favor of this
will be given below. Concerning the distribution of the eigenvalues of the
classical evolution operator, our second guess is therefore that
asymptotically (i.e., for resonances located far from the real axis) the
statistical properties of the imaginary part of the Ruelle--Pollicott
resonances located on the critical line ${\rm Re} (s) = h_{\rm top}/2$
coincide with those of the corresponding quantum (eigen) wavenumbers, which
are random matrix like generically. This second hypothesis could be seen as an
extension of the Bohigas--Giannoni--Schmit conjecture \cite{bgs} to the
statistical properties of the spectrum of the classical evolution operator.
This way, the random matrix properties in fully chaotic systems would have a
fully classical counterpart.

In semiclassical theories, the quantum correlations can be related to
correlations acting among the actions of the periodic orbits \cite{argaman}.
In scaling systems like billiards, the action coincides, up to a constant
factor, with the period (or length). Therefore action correlations are
equivalent to period correlations. Since, as we have here shown, the density
of periods of periodic orbits can be expressed in terms of the eigenvalues of
the classical evolution operator, it follows that the correlations between
periods of periodic orbits can be expressed in terms of correlations acting
among the Ruelle--Pollicott resonances. The quantum spectral correlations are
thus mapped, via semiclassics, into classical spectral correlations. Through
this connection, the RMT conjecture of the quantum fluctuations should have a
classical counterpart, that applies to the fluctuation properties of the
position in the complex plane of the Ruelle--Pollicott resonances. This gives
some support to our second hypothesis.

The two previous conjectures determine the gross features of the low lying
spectrum of the evolution operator in fully chaotic billiards and, as a
consequence, of the long time behavior of the density and correlations of
periodic orbits. The random matrix universality observed in quantum systems
may have, by semiclassical arguments, a classical counterpart. The ``ergodic''
zero located at the origin certainly plays an important role \cite{aasa}. We
are here suggesting a clear and explicit additional link between the
statistical properties of the classical and quantum spectrums, now involving
the complex resonances.

To conclude, we briefly discuss the spectrum of the evolution operator for
discrete maps, and show that important qualitative differences with respect to
smooth flows occur. We have in mind area--preserving classically chaotic maps
acting on two--dimensional phase spaces, like for example the kicked Harper or
the kicked top. The time now takes only discrete values, $\tau = n$, $n=1, 2,
3, \ldots$ (in some arbitrary units), and the set of possible lengths of
periods of periodic orbits is trivial, just integers. The ``return
probability'' or trace of the evolution operator is still expressed as a sum
over the periodic points \cite{book},
\begin{equation}\label{trmap}
R (n) = {\rm tr} {\cal L}^n = \sum_p \sum_{r=1}^\infty \frac{n_p}{\detp} 
\ \delta_{n, r n_p} = \sum_\gamma g_\gamma \ {\rm e}^{- \gamma n}  \ ,
\end{equation}
where $n_p$ is the (integer) period of the periodic orbit $p$, and the
$\gamma$'s are the Ruelle--Pollicott resonances. The latter are the zeros of
(\ref{zs}), making the replacement $\tp \rightarrow n_p$. Inverting
Eq.(\ref{trmap}) as in \S \ref{4}, a formula follows for the number of
primitive periodic orbits of period $n$,
\begin{equation}\label{Nn}
{\cal N} (n) = \frac{1}{n} \sum_{m/n} \mu (m) \ |\det (M_{n/m} - 1) | \
\sum_\gamma g_\gamma \ {\rm e}^{- \gamma n/m} \ .
\end{equation}
As in \S \ref{4}, we have ignored the fluctuations of the factor $|\det
(M_\ell - 1) |$ which may occur at short periods. An alternative formula,
equivalent to (\ref{nr2}), is obtained from the corresponding Ruelle zeta,
${\cal N} (n) = n^{-1} \sum_{m/n} \mu (m) \ \sum_\eta g_\eta \ {\rm e}^{\eta
n/m}$, where the $\eta$'s are the poles and zeros of Eq.(\ref{ztop}) with the
replacement mentioned above. Since the time $\tau = n$ changes by unit steps,
the smallest scale over which temporal variations can occur is one. It follows
from Eq.(\ref{trmap}) that the complex resonances satisfy $- \pi < {\rm Im}
(\gamma) \leq \pi$. Therefore, the natural variable to analyze discrete maps
is not $s$, but rather $z = {\rm e}^{-s}$. By this transformation, the ergodic
zero $\gamma_0 = 0$ is located at $z=1$, and other real and complex
Ruelle--Pollicott resonances with ${\rm Re} (\gamma) > 0$ lie inside the unit
disk $|z| \leq 1$.

Substantial differences are expected between the structure of the spectrum of
$Z (s)$ and $Z_R (s)$ for chaotic maps with respect to that of chaotic
continuous flows, and in particular with respect to chaotic billiards. The
main reason for that is the simplicity of the spectrum of periods of periodic
orbits in the case of maps. Since that spectrum is made of integers, the only
non--trivial information carried by Eqs.(\ref{trmap}) and (\ref{Nn}) concerns
variations in the stability factors and number of orbits of a given period
(e.g., average coarse--grained properties). This is in contrast with
continuous chaotic billiards, where a much more subtle and delicate
information is encoded in the spectrum of resonances, namely the nontrivial
distribution of the periods of the orbits. In maps, that distribution
collapses to a simple and highly degenerate spectrum. The simplicity of the
average properties encoded in Eq.(\ref{trmap}), without fine--grained
structure, implies a simpler spectrum of resonances. In particular, no
concentration of resonances over a ``critical'' line is expected to occur
(this would be a critical circle inside the unit disk in the $z$ variable),
since its presence is associated to a harmonic decomposition of the
distribution of periods in $R(\tau)$, whereas $R(n)$ strictly tends to a
constant for long times. In hyperbolic maps, isolated resonances, without any
special structure in the radial direction, are therefore expected generically
inside the unit circle. No particular connection with random matrix theory is
therefore established concerning the statistical properties of the resonances
of chaotic maps. This picture seems to be confirmed by recent numerical
simulations \cite{whbms}.

Semiclassically, the reason for the important differences with respect to
chaotic billiards is also clear. Random matrix requires correlations between
actions of periodic orbits. In scaling systems like billiards, the action of
an orbit is proportional to its length (or period). Therefore action
correlations translate into length (or period) correlations, which in turn,
through Eq.(\ref{rhor}), translate into resonance correlations. In maps, the
set of possible periods (which is trivial) is not related to that of actions
(which is nontrivial). Therefore, the spectrum of the evolution operator,
which is closely related to the spectrum of periods of the periodic orbits,
looses its connection with random matrix theory.



\begin{thebibliography}{99}
\bibitem{foot1} If the leading resonances would have satisfied $q_\gamma >
h_{\rm top}$, then the leading $\ti$ asymptotic oscillatory contributions to
the density would have been terms with the largest possible value of $m$ in
the sum. But this is inconsistent with the fact that repetitions do not
dominate the proliferation of periodic orbits.
\end{thebibliography}
\end{document}